\documentclass[manuscript]{aastex}

\shorttitle{KUV 01584-0939: A Helium-transferring Cataclysmic Variable with an Orbital Period of 10 Minutes}
\shortauthors{Brian Warner \& Patrick A. Woudt}

\begin{document}

\title{KUV 01584-0939: A Helium-transferring Cataclysmic Variable\\
 with an Orbital Period of 10 Minutes\altaffilmark{1}}

\author{Brian Warner and Patrick A. Woudt}
\affil{Department of Astronomy, University of Cape Town, Rondebosch, 7700
South Africa}
\email{warner@physci.uct.ac.za, pwoudt@artemisia.ast.uct.ac.za}

\altaffiltext{1}{This paper uses observations made from the South African Astronomical Observatory (SAAO).}

\begin{abstract}
High speed photometry of KUV 01584-0939 (alias Cet3) shows that it has a period
of 620.26~s. Combined with its hydrogen-deficient spectrum, this implies that it 
is an AM CVn star. The optical modulation is probably a superhump, in which
case the orbital period will be slightly shorter than what we have observed.
\end{abstract}

\keywords{cataclysmic variables, stars: binaries: close, star: individual (KUV 01584-0939)}

\section{INTRODUCTION}

The star KUV 01584-0939 (hereafter referred to as Cet3, as in the
Downes, Webbink \& Shara (1997) catalogue of Cataclysmic Variables (CVs))
was discovered during the Kiso survey (Kondo, Noguchi \& Maehara 1984) for ultraviolet rich objects.
A spectrum of Cet3 covering the wavelength range 4000 -- 7100 {\AA} was obtained
by Wegner, McMahan \& Boley (1987) who drew attention to the great strength
of He\,II emission and the weakness of the Balmer emission lines.
We have observed Cet3 as part of a high time
resolution photometric survey of faint CVs (Woudt \& Warner 2001).

\section{OBSERVATIONS AND RESULTS}

\begin{figure*}
\plotone{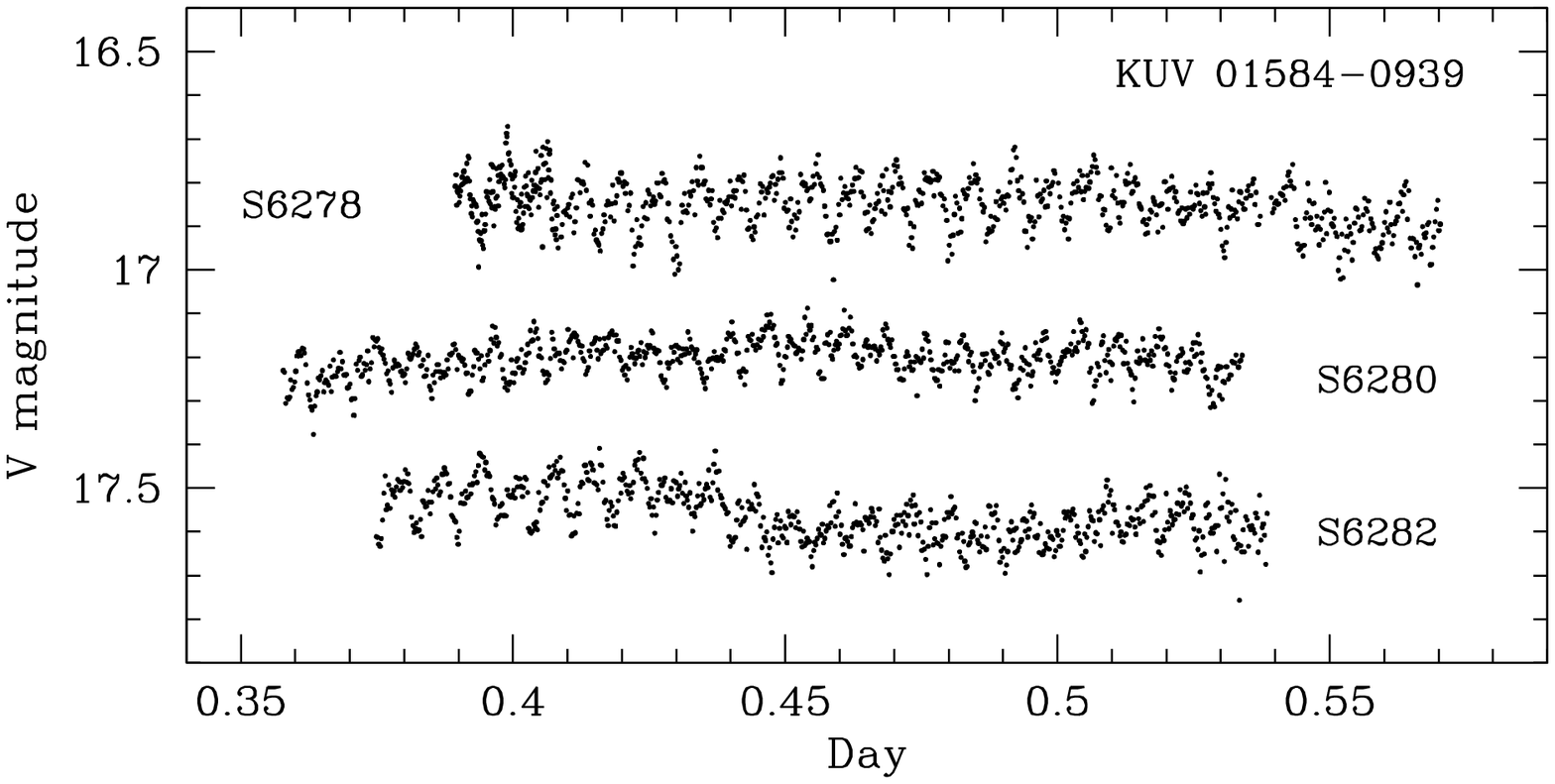}
\caption{The optical light curves of KUV 01584-0939 on 2001 October 18, 19, and 20. The upper
light curve is at the correct brightness. The others (in chronological order from top to bottom) have
been displaced vertically by 0.4 mag and 0.7 mag, respectively.}
\label{lccet3}
\end{figure*}

Our high speed photometric observations were made with the 
University of Cape Town CCD photometer (O'Donoghue 1995) mounted on the 74-in Radcliffe
reflector at the Sutherland site of the South African Astronomical Observatory on the
nights of 18 -- 21 October 2001. 
Cet3 is quite faint ($m_V \sim 16.9$) so we
used `white light', counting all photons detected by the CCD. The light curves
for the three long observing runs, obtained on consecutive nights, are shown in Figure~\ref{lccet3}.

\begin{figure}
\plotone{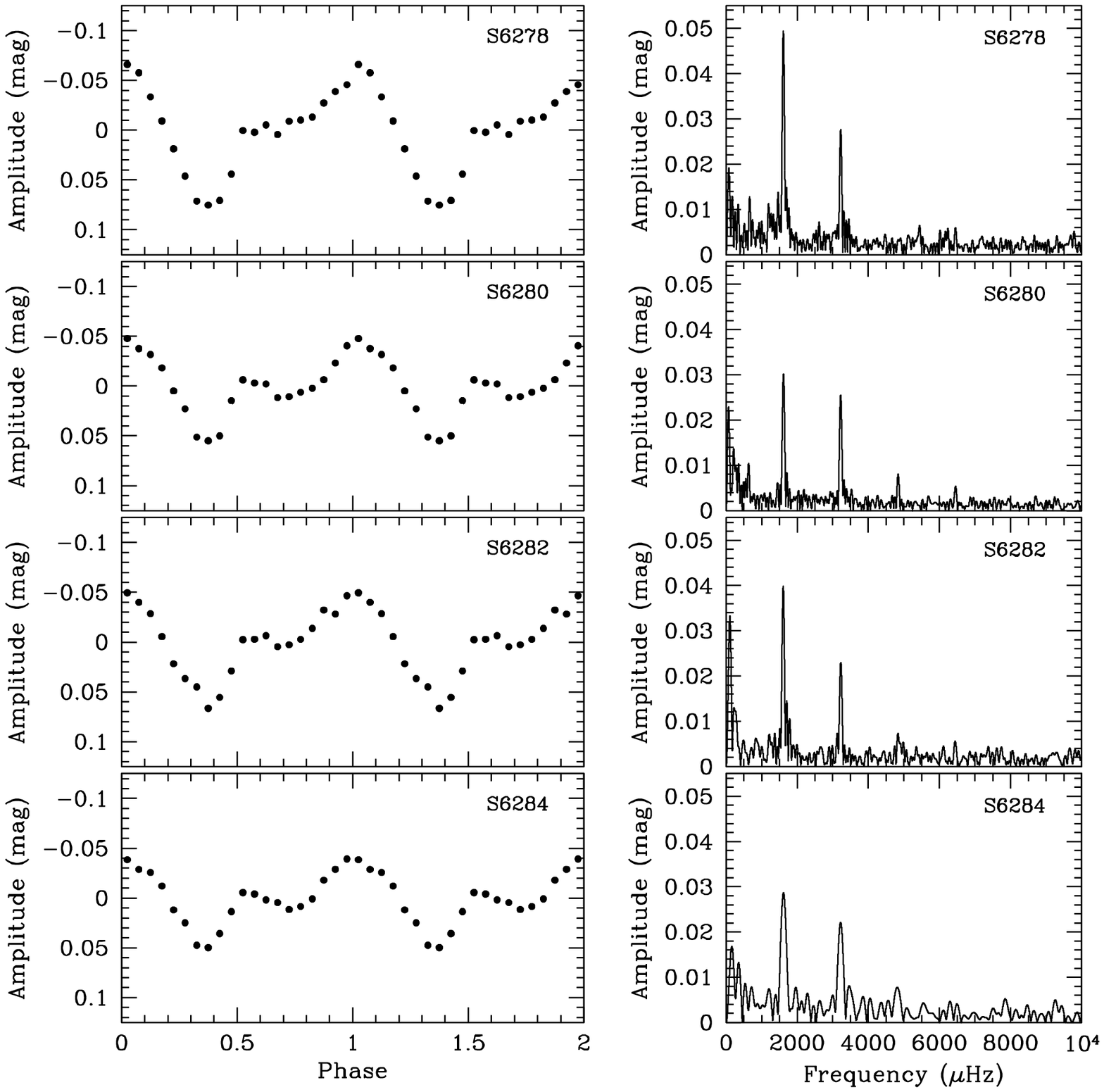}
\caption{The mean pulse profiles of the 620.26-s period (left panels) and 
Fourier transforms (right panels) for each of the individual observing runs.}
\label{ftcet3}
\end{figure}

From our observations we find that Cet3 is a variable star with a period of 620.26 $\pm$ 0.02 secs and 
a peak-to-peak range of $\sim$0.15 mag. The Fourier transform (FT) of our brightness measurements
shows a fundamental and its harmonics. 
The presence of harmonics
implies that the modulation is not sinusoidal. In Figure~\ref{ftcet3} we present a compact view
of the FTs and the mean pulse profiles for each of the 4 observing runs. The mean
pulse profiles are phased according to the ephemeris given in Eqn.~1.
There are only small
changes in average profile from night to night, but individual pulse profiles vary
as though there is some rapid flickering present.
In Figure~\ref{lccet3} it can also be seen that slow changes in mean brightness can also
occur. We have found no persistent features in the FTs other than the 620 s fundamental
and its harmonics. The ephemeris for maximum light of Cet3, derived from a non-linear
fit of sinusoids to the fundamental and the first two harmonics, is

$${\rm HJD_{max}} = 2452201.391401 + 0\fd0071788 (\pm 2) \, {\rm E}. \ \eqno{(1)}$$

Photometric periods of 10$^2$ -- 10$^3$ s in stellar sources can originate from a number
of different structures: including, e.g., orbital periods of low mass X-ray binaries (Bradt \& McClintock 1983)
containing a neutron star, or of the AM CVn stars mentioned above, or the rotation periods
of magnetic white dwarfs in CVs (known as intermediate polars (Warner 1995a)). We can exclude
a number of other possibilities (e.g., pulsations) because of the exceptional
nature of the spectrum of Cet3. The latter also helps us to select from among the orbital
or rotational options.

The most striking feature of the optical spectrum (Wegner et al.~1987) of Cet3 is the strength of the He\,II
lines at 4026 and 4686 {\AA}, and their presence at 4100, 4199 and 5412 {\AA}. The C\,IV emission
line at 5802 {\AA} and the very blue continuum emphasize that the source is very hot. 
These identifications raise the
possibility that the weak apparent Balmer emission is not from H\,I: it is more probably
the Pickering series of He\,II, in which case Cet3 is hydrogen deficient, probably belongs
to the AM CVn class of helium transferring interacting binaries, and the very blue continuum
is that of an accretion process. No low mass X-ray binary
or intermediate polar has a spectrum resembling that of Cet3. The slow changes in mean brightness
and the presence of flickering are characteristics of a mass-transferring close binary system.

\section{DISCUSSION}

AM CVn stars evolve from normal composition binaries which have experienced two phases of
mass exchange, exposing the helium cores of the original stars. During their orbital evolution
such systems pass through a minimum period near 4 mins and thereafter evolve to longer
periods (Tutukov \& Yungelson 1996). The driving mechanism for orbital evolution in such short period binaries
is loss of angular momentum through emission of gravitational radiation (Paczynski 1967). This
determines the rate of mass transfer, $\dot{M}$, and predicts (Warner 1995b) values $\sim 1 \times 10^{-8}$
M$_\odot$ y$^{-1}$ at $P_{orb}$ = 10 min, decreasing steeply to $4 \times 10^{-11}$ 
M$_\odot$ y$^{-1}$ at $P_{orb}$ = 40 min. These rates are compatible with the range of 
observed properties of the seven optically observed (Warner 1995b; Ruiz et al.~2001) AM CVn stars and the X-ray 
system (Cropper et al.~1998). The latter (V407 Vul = RX\,J1914+24) has $P_{orb}$ = 569 s, is thought to
be a polar (i.e., the primary is strongly magnetic, so its rotation is synchronised with the 
revolution of the secondary), but is very faint optically because it is obscured by
an interstellar cloud. In contrast, Cet3 has a similar period and is relatively 
accessible at $m_V = 16.9$.

If Cet3 has $\dot{M} \sim 1 \times 10^{-8}$ M$_\odot$ y$^{-1}$, its accretion luminosity
will be $\sim$ 5 L$_\odot$ for a 0.7 M$_\odot$ white dwarf primary. Most of this will be 
radiated at the inner boundary layer of the accretion disc at a temperature of
$\sim 3 \times 10^5$ K, so Cet3 should be a soft X-ray source of high intrinsic luminosity. 
This also accounts for the strength of the helium and carbon ionic spectra in Cet3. We calculate that d{$P_{orb}$}/d$t$
$\sim 6 \times 10^{-12}$, which implies that d{$P_{orb}$}/d$t$ should be measurable in
only a few years of observations if the modulation indeed has an orbital origin.
We have no means at present to estimate the distance of Cet3, but we expect it to be a
strong source of gravitational radiation (Warner 1995b; Hils \& Bender 2000).

We can estimate parameters for Cet3 in the following way. As the secondary must fill its Roche
lobe, a $P_{orb}$ of 620 s implies a mean density $\bar{\rho} \simeq 3.6 \times 10^3$ gm
cm$^{-3}$ (see equation 2.3b of Warner 1995a). From the mass-radius relation for low mass
white dwarfs (Tutukov \& Yungelson 1996), this gives a mass for the secondary $M(2)$ = 0.066 M$_\odot$.
For an assumed primary mass $M(1)$ = 0.7 M$_\odot$, we have $q = M(2)/M(1)$ = 0.094 and a binary
separation of $9.9 \times 10^9$ cm. Cet 3 is a very compact system: from standard
formulae (Warner 1995a) we find an accretion disc radius $\sim 5.4 \times 10^9$ cm, the radius $R(2)$
of the secondary $\sim 2.1 \times 10^9$ cm, and the radius of the primary $R(1) \sim 0.85
\times 10^9$ cm. The accretion disc therefore extends only $\sim 5.4 R(1)$ above the surface
of the primary and will be strongly irradiated by the $\sim 3 \times 10^5$ K source at the inner
edge of the disc.  In addition, for $\dot{M} \sim 1 \times 10^{-8}$ M$_\odot$ y$^{-1}$ the
maximum temperature in the disc itself is $\sim 6.2 \times 10^4$ K. The surface of the 
secondary intercepts $\sim$ 1\% of the soft X-ray emission and, with allowance for shielding
by the accretion disc, the side facing the 
primary will as a result be heated to $\sim 1.8 \times 10^4$ K.

These parameters are of use in attempting to understand the modulation profile 
in Cet3. For a very small mass-ratio ($q$ = 0.094) the accretion disc nearly fills
the Roche lobe of the primary (Warner 1995a) and partial eclipses of the disc are quite
probable. The shallow but wide minima in the Cet3 light curve could therefore be
eclipse features. The profile could alternatively be a combination of reflection
effects from the secondary and aspect effects of the disc and bright spot.
However, a small mass ratio $q$ and high $\dot{M}$ usually result
in perturbation of the accretion disc into an elliptical form with consequent precession (Warner 1995a).
The AM CVn stars with high $\dot{M}$ show `superhumps' with periods a few percent 
longer than $P_{orb}$, which result from the tidal stresses in a precessing disc (Patterson et al.~2001). It is probable,
therefore, that the modulation in Cet3 is a permanent superhump, in which case $P_{orb}$ will
be slightly shorter than the period we have measured. Alternatively, Cet3 may be a polar
like V407 Vul (the absence of two distinct periods disfavours an intermediate polar interpretation). 
It should be possible to distinguish
between these various interpretations when high time resolution spectroscopy is available.

\section*{ACKNOWLEDGEMENTS}

We thank Dr.~D.~Kilkenny for helpful comments. This research is supported by grants from
the University of Cape Town.

\begin{thebibliography}{}                                  

\bibitem[Bradt83]{b83}     Bradt, H.V.D., \& McClintock, J.E. 1983, \araa, 21, 13
\bibitem[Cropper98]{c98}   Cropper, M., Harrop-Allin, M.K., Mason, K.O., Mittaz, J.P.D., 
     Potter, S.B., \& Ramsay, G. 1998, \mnras, 293, L57
\bibitem[Downesea97]{d97}  Downes, R.A., Webbink, R.F., \& Shara, M.M. 1997, \pasp, 109, 345
\bibitem[Hils2000]{h00}    Hils, D., \& Bender, R.L. 2000, \apj, 537, 334
\bibitem[Kondo84]{k84}     Kondo, M., Noguchi, T., \& Maehara, H. 1984, Ann. Tokyo Astr. Obs., 20, 130
\bibitem[Odonoghue95]{o95} O'Donoghue, D. 1995, Baltic Astr., 4, 519
\bibitem[Paczynsky67]{p67} Paczynski, B. 1967, Acta Astr., 17, 287
\bibitem[Patterson01]{p01} Patterson, J., Fried, R., Rea, R., et al. 2001, \pasp, in press
\bibitem[Ruiz2001]{r01}    Ruiz, M.T., Rojo, P.M., Garay, G., \& Maza, J. 2001, \apj, 552, 679
\bibitem[Tutukov96]{t96}   Tutukov, A., \& Yungelson, L. 1996, \mnras, 280, 1035
\bibitem[Warner95a]{w95a}   Warner, B. 1995a, Cataclysmic Variable Stars (Cambridge: Cambridge University Press)
\bibitem[Warner95b]{w95b}   Warner, B. 1995b, \apss, 225, 249
\bibitem[Wegner87]{w87}    Wegner, G., McMahon, R.K., \& Boley, F.I. 1987, \aj, 94, 1271
\bibitem[Woudt2001]{w01}   Woudt, P.A., \& Warner, B. 2001, \mnras, 328, 159

\end{thebibliography}
\end{document}